\documentstyle[12pt]{article}
\newcommand{\gsim}{\mathrel{\hbox{\rlap{\lower.55ex \hbox {$\sim$}}
                   \kern-.3em \raise.4ex \hbox{$>$}}}}
\newcommand{\lsim}{\mathrel{\hbox{\rlap{\lower.55ex \hbox {$\sim$}}
                   \kern-.3em \raise.4ex \hbox{$<$}}}}
\newcommand {\vectwo}[2] {\left(\begin{array}{c}#1\\#2\end{array}\right)}

\newcommand {\beq} {\begin{equation}}
\newcommand {\eeq} {\end{equation}}

\newcommand {\ncl} {non-contractible loop}
\newcommand {\Ncl} {Non-contractible loop}
\newcommand {\ncls} {non-contractible loops}
\newcommand {\ncs} {non-contractible sphere}

\newcommand {\ncss} {non-contractible spheres}

\newcommand {\YMth} {Yang-Mills theory}
\newcommand {\YMHth} {Yang-Mills-Higgs theory}

\newcommand {\YMHths} {Yang-Mills-Higgs theories}
\newcommand {\sph} {sphaleron}

\newcommand {\cs} {configuration space}
\newcommand {\Sstar} {S$^{\star }$}
\newcommand {\Istar} {I$^{\star }$}
\newcommand {\Ibar}  {$\bar{{\rm I}}$}
\newcommand {\IIbar} {I$\, \bar{{\rm I}}$}

\newcommand {\fes} {forward elastic scattering amplitude}

\newcommand {\fess} {forward elastic scattering amplitudes}
\newcommand {\cme} {center-of-mass energy}

\newcommand {\dstar} {$d^{\star }$}
\newcommand {\Astar} {$A^{\star }$}
\newcommand {\Wstar} {$W^{\star }$}
\newcommand {\Phistar} {$\Phi^{\star }$}
\newcommand {\rhostar} {$\rho^{\star }$}
\newcommand {\ES} {E_{{\rm S}}}
\newcommand {\rmI} {{{\rm I}}}

\begin{document}

\begin{titlepage}
\hspace*{\fill}
NIKHEF-H/93-02
\newline
\hspace*{\fill}
March 1993
\begin{center}
	\vspace{2\baselineskip}
	{\Large \bf Existence of a new instanton in constrained \YMHth }\\
	\vspace{1\baselineskip}
\renewcommand{\thefootnote}{\fnsymbol{footnote}}
	{\large
	 F. R. Klinkhamer
\footnote{ ~E-mail address : klinkhamer @ nikhef.nl} \\
	 CHEAF/NIKHEF--H \\ Postbus 41882 \\
	 1009 DB Amsterdam \\ The Netherlands \\
	}
\renewcommand{\thefootnote}{\arabic{footnote}}
\setcounter{footnote}{0}
	\vspace{3\baselineskip}
	{\bf Abstract} \\
\end{center}
{\small \par
Our goal is to discover possible new 4-dimensional euclidean solutions
(instantons) in fundamental $SU(2)$ \YMHth,
with a constraint added to prevent collapse of the scale.
We show that, most likely, there exists one particular
new constrained instanton
(\Istar) with vanishing Pontryagin index. This is based on a topological
argument that involves the construction of a \ncl
{}~of 4-dimensional configurations with a certain  upperbound on the action,
which we establish numerically.
We expect \Istar ~to be the lowest action non-trivial solution in
the vacuum sector of the theory.
There also exists a related static, but unstable, solution,
the new sphaleron \Sstar.
Possible applications of \Istar ~to the electroweak interactions
include the asymptotics of perturbation theory
and the high-energy behaviour of the total cross-section.
}
\end{titlepage}

\section{Introduction}

We have conjectured \cite{K} the existence
of a new constrained instanton in the vacuum sector of euclidean
$SU(2)$ \YMHth. Our argument consists of two steps : \newline
(1) the construction of a \ncl ~ of 4-dimensional configurations
of the Yang-Mills and Higgs fields, starting and ending at the
classical vacuum; \newline
(2) the proof of a certain upperbound on the action over this \ncl,
which is crucial for having a genuine new solution, as will be
explained later on. \newline
For the last step, which is by far the most difficult of the two, we have
to resort to numerical methods. The numerical results of our previous
paper \cite{K} were not entirely conclusive in establishing this upperbound
on the actionprofile. Here, we present further numerical results
that {\em are} conclusive, at least for certain values of the
parameters in the theory.

Compared to our earlier work we have made the following improvements.
First, the constraint on the scale is treated dynamically, whereas before
the scale was fixed by hand.
Second, the ansatz of the \ncl ~is generalized, in order to give the
fields more freedom to relax to a lower value of the action.
Third, the efficiency and accuracy  of the numerical methods for solving the
variational equations from the ansatz were increased significantly.
The combined effect of these
three improvements allows us to establish the upperbound on the
actionprofile, which turns out to be a rather delicate affair.
Having established this upperbound does not rigorously prove the
existence of a new instanton, but makes it very likely
in our opinion. In that case, we also have, from our results for the \ncl,
an approximation of the exact solution \Istar.

The present paper is primarily concerned with the
existence of a new classical solution, and we refer the reader to \cite{K}
for the physics that motivates our search. The outline of the
present article is as follows.
In sect. 2 we discuss the constraint on the scale and what solutions
precisely we are after.
In sect. 3 we outline a general strategy for the search of new
solutions, which is to find \ncls ~in \cs.
One particular \ncl ~is presented in sect. 4.
First, we discuss, in subsect. 4.1, the basic ingredients of our construction;
then, we give, in subsect. 4.2, the ansatz in full detail.
The general behaviour of the action
over this \ncl ~is discussed in subsect. 4.3 .
The actual numerical results for the actionprofile
are presented in subsect. 5.2. These results show that, for small enough
values of the Higgs mass at least, the desired upperbound on the
actionprofile holds, which is the main result of this paper.
As a byproduct we obtain some numerical results for the
well-known BPSTH instanton \cite{BPST,H};
these results are given in subsect. 5.1 .
Finally, we return, in sect. 6, to the
possible existence of the \Istar ~solution
in $SU(2)$ \YMHth ~and compare this with the situation in other theories.
We also discuss, very briefly, potential applications
of \Istar ~to the physics of the electroweak interactions.
There are two appendices. In the first of these,
Appendix A, we give an outline of the calculations behind the
numerical results of sect. 5. In the second, Appendix B, we describe
a similar calculation for a static, 3-dimensional solution,
the new sphaleron \Sstar, which
is directly  related to the new instanton \Istar.
The reader who is not interested in the details is advised to
concentrate on sects. 2 and 3, possibly subsect. 4.1, and sect. 6.

\newpage
\section{Constrained Instantons}

We consider a \YMHth ~with $SU(2)$ gauge fields $W$ and a single, complex
Higgs doublet $\Phi$.
The euclidean action of this theory is
\beq
A_{\rm YMH} = \int_{{\rm R}^{4}} d^{4}x
 \left[ - \frac{1}{g^{2}}\; \frac{1}{2}\; {\rm Tr} \: W_{\mu \nu}^{2}
     + |D_{\mu} \Phi|^{2}
     + \lambda \left( |\Phi|^{2} - \frac{v^{2}}{2} \right)^{2} \: \right],
\label{eq:AYMH}
\eeq
where $W_{\mu \nu} \equiv \partial _{\mu} W_{\nu} - \partial _{\nu} W_{\mu}
+ [W_{\mu},W_{\nu}]\,$, $D_{\mu} \equiv \partial_{\mu} + W_{\mu}$ and
$W_{\mu} \equiv W_{\mu}^{a} \, \sigma^{a}/(2i)\,$,  with
$\sigma^{a}$ the standard Pauli matrices. Here, and in the following,
Greek indices run from 0 to 3 and Latin indices from 1 to 3.
The masses of the three $W$ vector bosons and the single Higgs scalar $H$
are $M_{W}=gv/2$ and $M_{H}=\sqrt{2\lambda}\, v$.
Finite action gauge field configurations can be classified by the
Pontryagin index (or topological charge), which takes on integer values
and is defined by
\beq
Q\equiv  \frac{1}{8 \pi^{2}} \int
         d^{4}x\;  \left[ - \frac{1}{2}\; {\rm Tr} \left(
         \frac{1}{2}\: \epsilon_{\kappa \lambda \mu \nu}\:
         W_{\kappa \lambda}\:  W_{\mu \nu} \right) \right] \: .
\label{eq:Q}
\eeq

A simple scaling argument shows the absence of non-singular 4-dimensional
classical solutions for the action (\ref{eq:AYMH}) :
any configuration can lower its action by collapse to a point.
For this reason one introduces a constraint on the scale and
later integrates over the corresponding collective coordinate $\rho$.
A convenient way to implement the constraint is to require \cite{A}
\beq
 \int d^{4}x \; {\rm O}_{ d} = 8\: \pi^{2}\: c\; \rho^{4-{d}}  \; ,
\eeq
where ${\rm O}_{d} = {\rm O}_{d}\, (W,\Phi)$  is a field operator
with canonical mass dimension $d > 4$. With a Lagrange multiplier
$\tilde{\kappa}$ one then looks for the stationary points of
\beq
 A_{\rm YMH} +
 \tilde{\kappa} \left( \rho^{{d}-4} \int d^{4}x \: {\rm O}_{d}
               - 8 \, \pi^{2} c \right)  \; .
\label{eq:Astationary}
\eeq

Concretely, we proceed as follows. First, we have to chose the constraint
operator ${\rm O}_{d}$. The actual choice is irrelevant in the end,
as long as we integrate, in the path integral, over the collective
coordinate $\rho$ (with the appropriate Jacobian). For purely technical
reasons, to be explained later on, we have chosen to work with the operator
\beq
 {\rm O}_{8}  \equiv \left( - \frac{1}{2} \; {\rm Tr} \;
                       \frac{1}{2} \: \epsilon_{\kappa \lambda \mu \nu} \:
                        W_{\kappa \lambda}\: W_{\mu \nu} \right)^2 \; .
\label{eq:O8}
\eeq
Having made this choice, we solve the field equations that result
from variations $\delta W$ and $\delta\Phi$ in the constrained action
\beq
 A \equiv  A_{\rm YMH} +
 \frac{\kappa}{g^{2}}\; M_{W}^{-4} \int d^{4}x \: {\rm O}_{\rm 8}\; ,
\label{eq:A}
\eeq
for a given value of the dimensionless coupling constant $\kappa$.
Let us denote the fields of one particular solution by
\Wstar ~and \Phistar, with a corresponding constrained action $A^{\star}$.
The scale of this solution is then determined by
\beq
 \rho^{\star} = \left[
                \frac{1}{8 \, \pi^{2} c} \;
                \int d^{4}x\;{\rm O}_{\rm 8}(W^{\star})
                \right]^{-1/4}
\label{eq:rhostar}
\eeq
and its Yang-Mills-Higgs action by
\beq
 A_{\rm YMH}^{\star} = A_{\rm YMH} (W^{\star},\Phi^{\star})  \; .
\label{eq:AYMHstar}
\eeq
These values $\rho^{\star}$ and $A_{\rm YMH}^{\star}$ are functions of
the coupling constant $\kappa$, as the fields
\Wstar ~and \Phistar ~are. Eliminating $\kappa$
one obtains $A_{\rm YMH}^{\star}$ as a function of $\rho^{\star}$.

This procedure is entirely straightforward. The
problem is to {\em discover\/}  the solutions of the field equations
from the constrained action (\ref{eq:A}).
Any such 4-dimensional euclidean solution will be called a
``constrained instanton''.
\footnote{~In pure \YMth ~``instantons'' are sometimes meant to refer
exclusively to solutions of the first-order self-duality
equations, which then solve the second-order field
equations by the Bianchi identity.
Here, we consider a (constrained) instanton to refer to
{\em any\/} localized, finite action solution of the general field equations
resulting from the euclidean (constrained) action.
          }
The prototypical constrained instanton is the solution of Belavin et al.
and 't Hooft \cite{BPST,H}.
We denote this basic instanton by ${\rm I}_{\rm BPSTH}$, or I for short.
Its  fields are given by (using differential forms)
\begin{eqnarray}
W &=& - f_{\rm ~I}(x) \;  dU_{\rm ~I} \; U_{\rm ~I}^{\, -1}   \nonumber \\
\Phi &=& h_{\rm ~I}(x)\; \frac{v}{\sqrt{2}}\; U_{\rm ~I}\vectwo{0}{1}\nonumber
\\
U_{\rm ~I} &=& \hat{x} \cdot \sigma \; ,
\label{eq:ansatzI}
\end{eqnarray}
with $x^{2} \equiv x_{\mu} x_{\mu}$, $\hat{x}_{\mu} \equiv x_{\mu} / x$ and
$\sigma_{\mu}\equiv (1,i \sigma_{m})$.
The instanton has Pontryagin index $Q_{\rm I}=1$.
By reflection there is also an anti-instanton \Ibar, with  equal action
but opposite topological charge $Q_{\bar{\rm I}}=-1$.
Only in the limit $\rho,\lambda \rightarrow 0$,  is it possible to find  an
analytical solution for the radial functions in the ansatz \cite{BPST,H}
\begin{eqnarray}
f_{\rmI} &=& \frac{x^{2}}{x^{2}+\rho^{2}} \nonumber \\
h_{\rmI} &=& \sqrt{f_{\rmI}} \; ,
\label{eq:fhI}
\end{eqnarray}
with a corresponding action
\beq
A_{\rm YMH}= \left( 1 + \frac{1}{2}\: (\rho M_{W})^{2}
             \right)
             \frac{8\pi^{2}}{g^{2}}  +
             {\rm O}\left( \lambda \right) \; .
\label{eq:AYMHI}
\eeq
In sect. 5.1 we will give some numerical results for the action at
finite values of $\rho$ and $\lambda$.

In this article we look for constrained instantons that are
not related to the BPST instanton ($ |Q| = 1$) \cite{BPST}
or other self-dual solutions ($ |Q| > 1$) \cite{ADHM} of pure \YMth.
In fact, our search is for solutions
in the {\em vacuum sector} ($Q=0$) of the constrained \YMHth
{}~(\ref{eq:A}). No such solutions are known at present.

\newpage
\section{Strategy}

Explicit construction of instantons in the vacuum sector of the constrained
theory is not feasible at present.
Instead, we present a topological argument for the possible existence
of at least one such solution. This topological argument involves the
construction of a suitable \ncl ~(NCL) of 4-dimensional configurations
of the fields $W$ and $\Phi$, starting and ending at the classical vacuum.
In addition, we may hope to gain some insight into the structure of
this solution, preliminary to an explicit construction of it.

The presence of non-contractible loops in \cs ~im\-plies,
as Taubes \cite{T} has shown in a somewhat different context,
the existence of new solutions of the classical field equations.
The intuitive idea is that by ``shrinking'' the NCL it gets ``stuck''
at a point in \cs, which corresponds (or is close) to a stationary
point of the action, i.e. a new solution of the field equations.
Specifically, this is a mini-max procedure, where we take the
maximum action on the NCL and try to minimize that action.
It is essential for this topological argument that the euclidean
actiondensity (\ref{eq:A}), with positive coupling constants
$\lambda$ and $\kappa$, is a positive semi-definite functional
of the fields, or, in other words, that it is bounded from below.
In mathematics this general subject is called Morse theory,
which aims to relate the critical points of a functional to the topology
of the function space on which the functional is defined.
The method of finding these stationary points by a mini-max
principle on \ncls ~goes under the name of Ljusternik-Snirelman theory.
References to the mathematical literature can be found in \cite{T}.

For the case at hand there is, however, one obvious loophole in the
argument. It could be, namely, that the mini-max procedure leads to an
approximate solution consisting of the BPSTH instanton solution I
and  anti-instanton \Ibar~at infinite separation.
In that case there would be no genuine new solution.
Clearly, this possibility is ruled
out if we are able to construct a particular NCL for which
\beq
{\rm max}\:  A_{{\rm NCL}}(\omega) < 2\:  A_{{\rm ~I}}\:  ,
\label{eq:Abound}
\eeq
where $\omega$ parametrizes the position along the loop.
This upperbound on the actionprofile is a necessary condition for
an existence proof of \Istar ~and the main goal of this article
is to establish it.

\newpage
\section{Non-contractible loop}

In this section we present an ansatz for a \ncl ~of configurations.
We start with the basic ingredients that go into the construction,
then give the details of the ansatz, and, finally, discuss the
expected behaviour for the action over the loop.
The actual profile of the action over the NCL has to be
determined numerically, these results will be presented in sect. 5.

\subsection{Basic ingredients}

The first step in the construction of the NCL is to give the structure
of the fields at infinity ($| x | \rightarrow \infty$). This will be based
on a non-trivial mapping
\beq
\tilde{U} : S_{3} \times S_{1} \rightarrow SU(2) \sim S_{3} \; ,
\eeq
where the first $S_{3}$ refers to the hypersphere at infinity, $S_{1}$
to the loop of configurations and $SU(2)$ to the gauge group.
The mapping should belong to the
non-trivial homotopy class in $\pi_{4}(S_{3})=Z_{2}$, so that, later on,
we have a loop of configurations that is indeed non-contractible.
A specific choice for the $SU(2)$ matrix $\tilde{U}$ is the following
\begin{eqnarray}
\tilde{U} &=&\exp [ (\omega+\pi/2)i\sigma_{3}]\;(\hat{x}\cdot \sigma)\;
             \exp [-(\omega+\pi/2)i\sigma_{3}]\;(\hat{x}\cdot \sigma)^{\dagger}
                                                        \nonumber \\
\hat{x}_{\mu} &\equiv & x_{\mu} / \: | x |              \nonumber \\
\sigma_{\mu}  &\equiv &(1,i \vec{\sigma}) \; ,
\label{eq:Utilde}
\end{eqnarray}
with a loopparameter $\omega \in [-\pi/2,+\pi/2]$.
The fields at infinity (pure gauge, of course) are given by
\begin{eqnarray}
\left. W \,\right]_{\infty}   &=& -  d\tilde{U} \; \tilde{U}^{-1} \nonumber \\
\left. \Phi\, \right]_{\infty}&=& \frac{v}{\sqrt{2}}\;\tilde{U}\vectwo{0}{1}\;.
\end{eqnarray}

The second step is to extend these fields inwards. For this purpose
we introduce two radial functions $\tilde{f}(x)$ and $\tilde{h}(x)$,
which approach 1 at infinity and vanish at the orgin, in order to
ensure continuity. Also, we extend the range of the loop parameter
$\omega$ to $[-\pi,+\pi]$, and make it into a real loop, starting and
ending at the {\em same} point,
i.e. the vacuum at $\omega=\pm \pi$.
In this way we arrive at the following NCL of configurations
\newpage
\begin{eqnarray}
\omega \in [-\pi,-\pi/2] \cup [+\pi/2,+\pi] \; :
                   &W   = & 0  \nonumber \\
                   &\Phi= &(\tilde{h} \, \sin ^{2} \omega +\cos ^{2}\omega)
                      \: \frac{v}{\sqrt{2}} \: \vectwo{0}{1}\nonumber  \\
                   &&\nonumber  \\
\omega \in [-\pi/2,+\pi/2] \; :
                   &W =& -\tilde{f}\:  d\tilde{U} \: \tilde{U}^{-1} \nonumber
\\
                   &\Phi=& \tilde{h}\:\frac{v}{\sqrt{2}} \; \tilde{U}
\vectwo{0}{1},
\label{eq:NCLsimple}
\end{eqnarray}
where the radial functions $\tilde{f}(x)$ and $\tilde{h}(x)$
have the boundary conditions
\begin{eqnarray}
\lim_{ |x| \rightarrow \infty} \tilde{f}, \,\tilde{h} &=& 1  \nonumber\\
\tilde{f}(0)     =\tilde{h}(0)                        &=& 0  \; .
\label{eq:fhsimple}
\end{eqnarray}
The precise form of these radial functions $\tilde{f}(x)$ and $\tilde{h}(x)$
is arbitrary in principle, but the mini-max idea is to optimize them
at the maximum point ($\omega=0$) on the NCL. This is done by solving
the variational equations for $\tilde{f}(x)$ and $\tilde{h}(x)$,
that result from inserting the ansatz (\ref{eq:NCLsimple}) with $\omega=0$
in the constrained action (\ref{eq:A}) and making
variations $\delta\tilde{f}$ and $\delta\tilde{h}$.

The NCL (\ref{eq:NCLsimple})
is quite elegant, but not good enough for our purpose.
The reason is that the actionprofile $A(\omega)$
has a maximum value $A(0)$ definitely greater than $2 A_{\rm ~I}$.
Hence, the inequality (\ref{eq:Abound}) does not hold
for the simplest possible NCL.
This brings us to the third, and final, step
in the construction of a suitable NCL. We start from the observation that
there are really two ``cores'' in (\ref{eq:Utilde}),
each of which resembles the (anti)instanton (\Ibar) I as given in
(\ref{eq:ansatzI}). The idea now is to separate these cores, in order
to profit from the attraction of the long-range fields present in the
theory, the Higgs field in our case.
This improved NCL, parametrized by $\omega \in [-3\pi/2,+3\pi/2]$,
has the following structure
\begin{tabbing}
Ia \hspace{.2cm} \= : \hspace{.4cm} \= $\omega \in [-3\pi/2,-\pi]$
   \hspace{.4cm} \= build up of an initial Higgs configuration;     \\
IIa\>:\>$\omega \in [-\pi,-\pi/2]$
    \> creation and separation of an \IIbar  ~pair;\\
III\>:\> $\omega  \in [-\pi/2,+\pi/2]$
    \> relative isospin rotation of the \IIbar ~pair;     \\
IIb\>:\> $\omega  \in [+\pi/2,+\pi]$
    \> collapse and annihilation of the \IIbar  ~pair; \\
Ib \>:\> $\omega  \in [+\pi,+3\pi/2]$
    \> clean up of the remaining Higgs configuration.
\end{tabbing}
The explicit construction of the NCL is somewhat involved and
will given in the next subsection.

\newpage
\subsection{Ansatz}

The configurations of the NCL are given by
\begin{eqnarray}
{\rm I~~~~~~}&:&W  =  0   \nonumber\\
            &&\Phi= (h \, \cos ^{2} \omega +\sin ^{2}\omega)
                        \: \frac{v}{\sqrt{2}} \: \vectwo{0}{1}  \nonumber\\
            &&  \nonumber\\
{\rm II,III}&:&W  = -f\:  dU \: U^{-1}   \nonumber\\
            &&\Phi= h  \: \frac{v}{\sqrt{2}} \: U \vectwo{0}{1},
\label{eq:NCL}
\end{eqnarray}
with the following $SU(2)$ matrix
\beq
U = \exp [ \Omega i\sigma_{3}]\;(\hat{y}_{-} \cdot \sigma)\;
    \exp [-\Omega i\sigma_{3}]\;(\hat{y}_{+} \cdot \sigma)^{\dagger}
\label{eq:U}
\eeq
and notation
\[
\begin{array}{rclrcl}
\hat{y}_{\mu\pm}&\equiv & y_{\mu\pm} / \: | y_{\pm}| & \hspace{1cm}
\sigma_{\mu}    &\equiv &(1,i \:\sigma_{m})\\
y_{m \pm}       &\equiv & g_{\pm}\: x_{m} &
y_{0 \pm}       &\equiv & x_{0 \pm} \equiv  x_{0} \pm D/2 \\
x_{\pm}^{2}     &\equiv & x_{0 \pm}^{2} + r^{2}  &
r^{2}           &\equiv & x_{m}^{2} \; \equiv \rho^{2} + z^{2}  \\
z               &\equiv & x_{3} &
t               &\equiv & x_{0}\; .
\end {array}
\]
Here $D=D(\omega)$ determines the core distance and $\Omega=\Omega(\omega)$
their relative isospin rotation
\beq
\begin{array}{lclll}
{\rm  Ia}&:& \omega \in [-3\pi/2,-\pi]&D      = 0
                                      &\Omega = 0  \\
{\rm IIa}&:& \omega \in [-\pi,-\pi/2]  &D     = d_{\rm max} \sin^{2}\omega
                                      &\Omega = 0  \\
{\rm III}&:& \omega \in [-\pi/2,+\pi/2]&D     = d_{\rm max}
                                      &\Omega = \omega + \pi /2  \\
{\rm IIb}&:& \omega \in [+\pi/2,+\pi]   &D    = d_{\rm max} \sin^{2}\omega
                                      &\Omega = \pi  \\
{\rm  Ib}&:& \omega \in [+\pi,+3\pi/2]  &D    = 0
                                      &\Omega = \pi  \; .
\label{eq:DO}
\end{array}
\eeq
The functions $f$, $h$ and $g_{\pm}$ in (\ref{eq:NCL}, \ref{eq:U})
are taken to be axial functions $f(r,t)$, $h(r,t)$ and $g_{\pm}(r,t)$,
with the following boundary conditions
\begin{eqnarray}
 \lim_{|x| \rightarrow \infty} f,\, h,\, g_{\pm} &=& 1 \nonumber\\
f(0,\pm D/2)=h(0,\pm D/2)                        &=& 0
\label{eq:bcs}
\end{eqnarray}
and reflection symmetry
\begin{eqnarray}
f(r,t)        &=& f(r,-t)  \nonumber \\
h(r,t)        &=& h(r,-t)  \nonumber \\
g_{\pm}(r,t)  &=& g_{\mp}(r,-t) \; .
\label{eq:refl}
\end{eqnarray}
The Pontryagin index (\ref{eq:Q})  vanishes for all configurations of the NCL,
precisely because of this reflection symmetry of the ansatz.
This completes the basic construction of the NCL. We remark that the only
difference compared to our previous paper \cite{K} is the
presence of the functions $g_{\pm}$ in the matrix $U$  (\ref{eq:U}),
but this change will turn out to be essential.

It remains to specify the four axial functions
$f(r,t)$, $h(r,t)$ and $g_{\pm}(r,t)$  that enter the ansatz.
Just as for the simple NCL of the previous subsection, the mini-max
procedure would be to insert the ansatz (\ref{eq:NCL}),
for $\omega=0$ and $D=d_{\rm max}$, into the action (\ref{eq:A})
and solve the variational equations for $f,h$ and $g_{\pm}$.
This turns out to be prohibatively difficult and, instead,
we make an explicit choice for $g_{\pm}$.
As will be explained in the next subsection, we want the cores
to become independent for large values of their separation $D$.
This can be achieved by the following choice, for example,
\beq
g_{\pm} = \frac
{ (x_{\mp}^{2}D^{2}/4)^{\alpha}+(\beta\rho_{\rm ~I})^{4\alpha}                }
{ (x_{\mp}^{2}D^{2}/4)^{\alpha}+(\beta\rho_{\rm ~I})^{4\alpha}+(D/2)^{4\alpha}}
\label{eq:gpm}
\eeq
with $x_{\pm}^{2}$ defined below (\ref{eq:U}) and
the parameter values $\alpha=4.0$ and $\beta=1.5$, obtained by trial and error.
Here $\rho_{\rm ~I} = \rho_{\rm ~I} (\kappa)$
is the scale for the BPSTH instanton I, see sect. 5.1 below.
Having fixed the functions $g_{\pm}$, we can solve numerically the variational
equations for $f$ and $h$. We will do this for both $\omega=0$ and $\pi/2$,
and for arbitrary values of the distance $D$. These solutions will be called
$\bar{f}(\omega,D)$ and $\bar{h}(\omega,D)$, where
the dependence on the spatial coordinates $r$ and $t$ is implicit.
In order to keep the variational equations as small as possible,
we have chosen the constraint operator (\ref{eq:O8}), and not, for example,
\[ \left( - \frac{1}{2}\; {\rm Tr} \; W_{\mu \nu} W_{\mu\nu}\right)^2 \: , \]
which would give significantly larger expressions.
Furthermore, we set the constant $c$ in the definition of the scale
(\ref{eq:rhostar}) to the value
\beq
 c \equiv 288 / 21 \; ,
\label{eq:c}
\eeq
in order to match the scale of the BPSTH instanton, which will be
determined in sect. 5.1.

With the solutions $\bar{f}(\omega,D)$ and $\bar{h}(\omega,D)$
in hand, we can, at last, specify the functions $f$ and $h$ that enter
the ansatz (\ref{eq:NCL}) for the NCL
\begin{eqnarray}
{\rm I,II} &:  &f=  \bar{f}(\pi/2,D) \nonumber  \\
            &  &h=  \bar{h}(\pi/2,D) \nonumber  \\
            &  &\nonumber  \\
{\rm III~} &:  &f= \bar{f}(0,    d_{\rm max}) \: \cos^{2}\omega +
                    \bar{f}(\pi/2,d_{\rm max}) \: \sin^{2}\omega  \nonumber\\
            &  &h= \bar{h}(0,    d_{\rm max}) \: \cos^{2}\omega +
                    \bar{h}(\pi/2,d_{\rm max}) \: \sin^{2}\omega \; ,
\label{eq:fhNCL}
\end{eqnarray}
with the distance $D=D(\omega)$ given by
(\ref{eq:DO}). Actually, we only need the $\omega=\pi/2$ solutions
$\bar{f}$ and $\bar{h}$ down to
some small value $D=d_{\rm min}$. We can then close the loop with
a simple deformation of these functions  and keep them non-singular, as
explained in our previous paper \cite{K} and in Appendix B at the
end of this one. By choosing $d_{\rm min}$ small enough, the action
can be kept arbitrarily small.

To summarize, the field configurations of the NCL are given by
(\ref{eq:NCL} -- \ref{eq:DO})
with the axial functions (\ref{eq:gpm}, \ref{eq:fhNCL}),
where $\bar{f}$ and $\bar{h}$ are the solutions of the variational equations
for $\omega=0$ or $\pi/2$ and $D \leq d_{\rm max}$.

\subsection{Actionprofile}

Our main interest lies in the actionprofile over the NCL. Here, we will
give a general discussion of the possible behaviour, in order to
prepare the way for the numerical results to be presented in
sect. 5.

The profile of the constrained action
over the NCL is essentially determined by the
behaviour of the constrained action $A(\omega,D)$
for the solutions $\bar{f}$ and $\bar{h}$ at $\omega=0$ or $\pi/2$.
In fact, $\bar{f}$ and $\bar{h}$ solve the variational equations,
so that the action attains its lowest value, within the ansatz,
precisely for these functions.
On physical grounds, we expect the following behaviour for $A$.
For $D \rightarrow \infty$ the constrained action should approach
twice that of the BPSTH instanton I,
whereas for $D \rightarrow 0$ the effects of
the Yang-Mills interactions should become important,
which can be either repulsive ($\omega=0$) or attractive ($\omega=\pi/2$).
The Yang-Mills interactions for large values of $D$  get suppressed
exponentially, with a length scale set by $M_{W}^{-1}$.
In this paper, we consider the case of vanishing
Higgs mass $M_{H}=0$ or $\lambda=0$.
This means that at very large distances $D$, where the Yang-Mills interactions
drop out, only the effect from the Higgs field remains. Moreover,
the Higgs interaction at large distances is attractive  and basically
independent of the relative isospin rotation $\Omega$ of the cores.
So, the simplest behaviour we expect is the one sketched in fig. 1a.
At this moment, we can also explain the need for the additional functions
$g_{\pm}$ in the ansatz (\ref{eq:U}). For $g_{\pm}=1$, namely,
there would be ``tidal effects'' from one core on the other, which
decrease rather slowly with the distance $D$. If, instead, $g_{+}$ vanishes
approximately near the $t=+D/2$, and $g_{-}$ near $t=-D/2$, the cores
become independent faster with increasing distance.
This, then, is the reason behind our choice
(\ref{eq:gpm}) for these functions.

As mentioned before,
the actionprofile over the NCL is essentially determined by the
behaviour of $A(0,D)$ and $A(\pi/2,D)$.
This follows from, in particular, (\ref{eq:DO}) and (\ref{eq:fhNCL}) above.
Concretely, the actionprofile,
which  is an even function of $\omega$, is obtained as follows.
We start in fig. 1a on the $\omega=\pi/2$ curve
at $D=0$ and move out to $D=d_{\rm max}$, then we go straight up
to the $\omega=0$ curve and back again, and finally we return
in the same way to $D=0$. We have verified that the action
over part III of the NCL, with the functions (\ref{eq:fhNCL}),
has indeed a single maximum at $\omega=0$.
{}From the curves of fig. 1a, then,  we would
conclude that the maximum action over the NCL can remain below
$2 A_{\rm ~I}$, provided we have a large enough value of $d_{\rm max}$.
In fact, the optimal choice (i.e. lowest possible maximum action)
would be to move out to a distance \dstar ~with a corresponding
maximum action \Astar, both of which are indicated in fig. 1a.
However, another possible behaviour of $A(0,D)$ and $A(\pi/2,D)$
is sketched in fig. 1b.
In that case, the maximum action would always stay above $2 A_{\rm ~I}$,
regardless of the value of $d_{\rm max}$.
A priori, there is no way to decide between the two types of behaviour
shown in fig. 1 and we need an explicit calculation to settle the matter.

\newpage
\section{Numerical results}

We present here some numerical results for the NCL constructed in the previous
section. These results are for the case of vanishing quartic
Higgs coupling constant $\lambda=0$. We also give some results for the
BPSTH instanton I.
Distances will be expressed in units of
$M_{W}^{-1}$ and the action in units of $8 \pi^{2} / g^{2}$, which is
the action of the BPST instanton in pure \YMth.
A brief outline of the numerical methods is given in Appendix A.

\subsection{BPSTH instanton}

We have solved numerically the variational equations for the radial
functions  $f_{\rm ~I}(x)$ and $h_{\rm ~I}(x)$,
that result from inserting the BPSTH ansatz (\ref{eq:ansatzI}) into the
constrained action (\ref{eq:A})
and making variations $\delta f_{\rm ~I}$ and $\delta h_{\rm ~I}$.
As explained in sect. 2,
these solutions $\bar{f}_{\rm ~I}$ and $\bar{h}_{\rm ~I}$
depend on the coupling constant $\kappa$ of the constrained
action $A$. We determine the scale $\rho$ of the instanton from
(\ref{eq:rhostar}), with the constant $c=144/21$,
and the Yang-Mills-Higgs action $A_{\rm YMH}$ from
(\ref{eq:AYMHstar}), where \Wstar ~and \Phistar ~now refer to the  ansatz
(\ref{eq:ansatzI}) with the functions $\bar{f}_{\rm ~I}$ and $\bar{h}_{\rm
{}~I}$.
This particular choice for the numerical value of the constant $c$
in (\ref{eq:rhostar}) reproduces the scale $\rho$ that enters
the analytical solution (\ref{eq:fhI}) of pure \YMth.
Our numerical results are collected in Table 1, where, for future reference,
we also give results for
$\lambda/g^{2} =1/8$ ($M_{H}=M_{W}$) and
$\lambda/g^{2} =100/8$ ($M_{H}=10 \, M_{W}$).
The $\lambda=0$ results for $A_{\rm YMH}$ vs. $\rho$
are compared, in fig. 2, with the expression (\ref{eq:AYMHI}),
which is valid in the limit $\rho \rightarrow 0$.
For values $\rho \gsim M_{W}^{-1}$ the radial functions of the numerical
solution differ significantly from the analytical functions (\ref{eq:fhI}),
resulting in a lower value of the action.
Another consequence is that, for finite values of the scale $\rho$,
the gauge fields are no longer self-dual.

The numerical results for the instanton I are relatively easy to
obtain, as the variational equations for
$\bar{f}_{\rm ~I}$ and $\bar{h}_{\rm ~I}$
are ordinary differential equations (ODEs). These results will serve
as a check on those of the NCL, to which we turn now.

\subsection{\Ncl}

We have solved numerically the variational equations for the axial
functions  $f(r,t)$ and $h(r,t)$,
that result from inserting the ansatz of the NCL (\ref{eq:NCL}),
for $\omega=0$ or $\pi/2$, into the constrained action (\ref{eq:A})
and making variations $\delta f$ and $\delta h$.
These variational equations consist of two coupled, non-linear
partial differential equations (PDEs), whose rather complicated structure
is explained in Appendix A.
For the numerical solutions $\bar{f}$ and $\bar{h}$,
we also determine the Yang-Mills-Higgs action  $A_{\rm YMH}$ and
the scale $\rho$ from the expressions (\ref{eq:AYMH}) and (\ref{eq:rhostar}),
respectively, with the numerical value (\ref{eq:c}) for the constant $c$
and the NCL configurations for \Wstar ~and \Phistar.
As discussed in subsect. 4.3, we are
primarily interested in the behaviour of the constrained action of the
solution as a function of the core distance $D$.
For $D \rightarrow \infty$ we reproduce the results of the previous subsection
for $2 A_{\rm ~I}$ (Table 1).
In fig. 3 we give the numerical results for the constrained action, normalized
to its asymptotic value, for a relatively large value of the constraint
coupling constant, namely $\kappa = 1$.
Figure 4 shows the same results on an expanded scale.
These numerical results seem to agree, for $\kappa = 1$,
with the simple behaviour sketched in fig. 1a.
In contrast, the results for $\kappa = 10^{-3}$,
shown in fig. 5, display the alternative behaviour of fig. 1b.
This different behaviour for small and large values  of the scale was also
seen in our previous results \cite{K}, but we cannot directly compare
the scale $\bar{\rho}$ there with the scale $\rho$ here.

We are reasonably confident that the results of figs. 3--5 are reliable,
for the following reasons.  First, the results are stable
when the calculation is repeated on different
grids (ranging from $25 \times 50$ to $100 \times 200$ points) and
with different compactified coordinates for $r$ and $t$,
see Appendix A.  The error of the data points in fig. 4 is estimated
to be less than $0.1 \%$, which is about a factor of 10 better than
the numerical accuracy of our previous results \cite{K}.
Second, the variational equations for $f$ and $h$
are solved by relaxation, which implies that the exact solution
can have a somewhat lower action. In other words, the ``dip''
of the $\omega=0$ curve in
fig. 4 can only get deeper. Third, and as is often the
case with large numerical calculations, we beleive our results
because they behave in the way we expect them to do.
In particular, the dip of fig. 4 becomes more shallow with
$\lambda/g^{2}$ increasing.
Also, we have an heuristic understanding of the origin,
for small values of $\kappa$,
of the ``bump'' on the $\omega=\pi/2$ curve in fig. 5.
Fourth, a similar calculation for static
fields gives comparable results, see Appendix B.

To summarize, we have constructed a NCL, for which the important upperbound
(\ref{eq:Abound}) holds, provided the scale $\rho$ is large enough.
Specifically, we obtained for $\lambda/g^{2}=0$ and $\kappa = 1$ (see figs.
3-4)
\beq
{\rm max}\:  A_{{\rm NCL}}(\omega)= 1.994 \: A_{\rm ~I} < 2\:  A_{{\rm ~I}}\:
{}.
\label{eq:Aboundresult}
\eeq
The optimal maximum ($\omega=0$) configuration of the NCL gives, moreover,
an approximation of the conjectured constrained instanton \Istar.
This configuration leads to the following estimates for \Istar
\begin{eqnarray}
 \rho^{\star}         &\sim&  1.9   \: M_{W}^{-1}  \nonumber \\
 d^{\star}            &\sim&  10 \; \: M_{W}^{-1}  \nonumber \\
 A_{\rm YMH}^{\star}  &\sim&  4.2\; \: 8 \pi^{2}  / g^{2}  \; .
\label{eq:estimates}
\end{eqnarray}
In fig. 6 we show the corresponding actiondensity
$\tilde{a}_{\rm YMH}(r,t)$ and Pontryagin-density $\tilde{q}(r,t)$,
averaged over the polar angle $\theta$, see Appendix A.
This shows that our configuration is still a very loose molecule
and we expect the exact solution \Istar ~to be tighter and more
cigar-like perhaps.
But it is also clear, from fig. 4 especially, that the Yang-Mills
cores are very hard and that \dstar, which is in essence the distance between
the points of vanishing Higgs field, cannot be much smaller than
the width ($\sim$ 2 \rhostar) of the configuration.

\newpage
\section{Discussion}

We have constructed in this paper a \ncl ~(NCL) with a maximum
constrained action less than twice that of the BPSTH instanton,
provided the length scale is fixed at a large enough value.
This was established for the case of vanishing Higgs mass $M_{H}=0$.
We expect it to be possible to extend the result to all values $M_{H} < M_{W}$.
Note that in the full electroweak theory there is
also the photon field, which can provide the necessary
attraction if the Higgs field becomes too short of range.
As it stands, this upperbound on the actionprofile over
the NCL is only a necessary condition in an eventual existence
proof. Still, we are optimistic about the existence of \Istar ~in $SU(2)$
\YMHth.  There exist, of course, analogous {\em static} solutions (sphalerons)
in \YMHth ~\cite{T,DHN,KM}, but there are also encouraging results
on {\em instantons} in pure \YMth, which we will now discuss.

There has been a long-standing conjecture \cite{AJ},
based on the analogy with harmonic maps from $S_{2}$ to $S_{2}$,
 that all solutions in
euclidean $SU(2)$ \YMth ~over $S_{4}$ are necessarily self-dual
or anti--self-dual
($ W_{\mu \nu}= \pm \frac{1}{2} \epsilon_{\mu \nu \kappa \lambda }
   W_{\kappa \lambda}$).
In the vacuum sector, in particular, there would be no other solutions
besides the classical vacuum itself.
This conjecture has been proven false recently.
In fact, Sibner et al. \cite{SSU} showed the existence of infinitely
many non--self-dual solutions in the vacuum sector, with an action
\beq
A_{\rm YM} =  m \: 16\, \pi^{2} / g^{2} + \Delta A_{\rm YM}\;,
\label{eq:AYMbound}
\eeq
\noindent
for integers $m \geq 2$ and $\Delta A_{\rm YM} <0$, which, most likely,
depends on $m$ also.
Later, several solutions were constructed
explicitely by Sadun and Segert \cite{SS}. The existence proof of
Sibner et al. \cite{SSU} goes by a mini-max procedure over \ncls,
where an important
ingredient is the inequality contained in (\ref{eq:AYMbound}),
which is analogous to ours (\ref{eq:Abound}).
The other main ingredient is an equivariant weak compactness theorem,
where equivariant refers to an $U(1)$ symmetry of their ansatz.
The integer $m$, which appears in (\ref{eq:AYMbound}), labels the
embedding in $SU(2)$ of this $U(1)$ symmetry
(rotation angle $\gamma \in [0,2\pi]$), namely by the matrix
$\exp [m \gamma  \sigma_{3} / (2 i) ]$ for the gauge field transformations.

Equivariance is known to be a powerful tool in Morse theory.
For this reason, it may be of importance to note that our ansatz
(\ref{eq:NCL}, \ref{eq:U}) also has an $U(1)$ symmetry, and precisely for
the case $m=1$ excluded in the existence proof \cite{SSU}
and the explicit construction \cite{SS} of the pure Yang-Mills solutions.
The difference is that our theory has an additional field,
the Higgs field namely, to provide the attraction necessary for
the crucial upperbound (\ref{eq:Abound}) on the action.
The constrained action (\ref{eq:A}) of this theory now
has three terms extra, compared to the case of pure \YMth.
The constraint term, in particular, is designed to prevent collapse
and, naively,  we see no way how a regular solution \Istar, related to the
presence of \ncls ~in configurations space, could fail to exist.

Just as the sphaleron S \cite{DHN,KM} is associated with
the BPSTH instanton I (loosely speaking S is a constant time slice of I),
we expect a new sphaleron \Sstar ~\cite{K-Sstar} to be associated with
\Istar. In Appendix B we give some numerical results for a non-contractible
{\em sphere} of 3-dimensional configurations, which support the existence
of this new sphaleron \Sstar.
After these results were obtained, we succeeded in constructing
the solution \Sstar, on which we will report elsewhere. We expect the
construction of \Istar ~to proceed in the same way, only with greater
technical complications. Henceforth, we take for granted the
existence of the new constrained instanton \Istar ~in $SU(2)$ \YMHth.

An interesting question is to see what happens when
massless fermions are introduced into the theory, or possibly fermions
with Yukawa couplings to the Higgs.
We then expect \Istar ~(and \Sstar) to have fermion zero-modes.
The reason is that, as explained in our previous paper \cite{K},
there is spectral flow of the Dirac eigenvalues along the NCL (\ref{eq:NCL}).
This spectral flow was indirectly monitored in
the numerical calculations of sect. 5
and our results support the claim that \Istar ~has fermion zero-modes.
Consequently, there should be a new ($B+L$ conserving)
effective fermion vertex from \Istar,
with double the number of lines compared to the one from the
BPSTH instanton \cite{H}.
We will now turn to different instanton solutions, which
may be related to \Istar ~in one way or another.

The pure $SU(2)$ Yang-Mills solutions \cite{SSU,SS} discussed above
can be expected to have counterparts in $SU(2)$ \YMHth,
i.e. there will be corresponding constrained instantons.
In general, these solutions will have a rather large action,
since for them we have already that $A_{\rm YM} \gsim 32 \: \pi^{2} / g^{2}$,
whereas $A_{\rm YM} \sim 16 \: \pi^{2} / g^{2}$ for \Istar.
We expect \Istar ~to be the lowest action non-trivial solution in
the vacuum sector of the theory.
More interesting, perhaps, is the inverse question, wether or not
the constrained instanton \Istar ~of $SU(2)$ \YMHth
{}~has a corresponding solution in pure \YMth.
We conjecture there to be no such counterpart in pure
\YMth, because we see no obvious source of attraction.
The fact that we found the upperbound on the action (\ref{eq:Abound})
to be violated for small values of the constraint coupling
constant (see fig. 5) is suggestive, but is not really conclusive, since it
could also mean that our NCL is not good enough.

Different \YMHths ~can also be considered. If \Wstar ~and \Phistar
{}~are the fields of the \Istar ~solution in $SU(2)$ \YMHth,
then it is possible to embed them into a larger theory,
provided that $SU(2)$ is a subgroup of the gauge group $G$ and that
the $SU(2)$ Higgs doublet (together with its vacuum expectation value)
can be embedded in the larger Higgs representation.
An example is $SU(3)$ \YMHth ~with a complex triplet of Higgs. The
embedding is then given by
\beq
 W =       \left(    \begin{array}{ccc}
                      0   &   0    &    0    \\
                      0   & \times & \times  \\
                      0   & \times & \times
                      \end{array}
             \right)                           \hspace{1cm}
\Phi =       \left(   \begin{array}{c}
                      0       \\
                      \times  \\
                      \times
                      \end{array}
            \right)                            \hspace{0.5cm} ,
\eeq
where the crosses indicate, symbolically, the $SU(2)$ solution.
Note that these embeddings can lead to
unexpected solutions of the field equations, since the larger theory
may not even have \ncls ~($\pi_{4}(G)=1$, as is the case
for $G=SU(3)$, for example). The NCL was a tool to find the
$SU(2)$ solution, but once we have found the solution, we can
forget about the NCL and simply verify
the fact that the ansatz solves the field equations, which is then
carried over to the larger theory.

Finally, we comment on possible applications of the conjectured new
instanton solution \Istar. Of course,  $SU(2)$ \YMHth ~is
at the heart of the Glashow-Weinberg-Salam model for the
electroweak interactions.
Evaluating euclidean path integrals for electroweak processes,
it may be important to know
that in addition to the classical vacuum there is a new stationary
point, i.e. the constrained instanton \Istar.
It is well-known \cite{L,BF} that such stationary points
(and vacuum instability,
in general) can play a role in the asymptotics of perturbation theory.
Furthermore, this new stationary point could contribute directly to the
euclidean path integrals of certain \fess, which control
the total cross-sections for the
corresponding processes. In our previous paper \cite{K} we have
shown that this contribution, evaluated classically
\footnote{
{}~For a semiclassical calculation we also need to know the negative modes
around the classical solution. In pure $SU(2)$ \YMth ~it has been shown
\cite{T-stab} that there must be at least two negative modes, instead of one;
wether or not the same holds in \YMHth ~is not known at the moment.
Note that even a  new, ``non-perturbative'' contribution to the real part only
of the \fes ~should have consequences, because of analyticity.
         },
suddenly becomes important as the parton \cme ~$\sqrt{s}$ increases.
In fact, this threshold energy is determined
by the structure of the \Istar ~solution, and from our
approximation of that solution we obtain
\beq
 (\sqrt{s})_{{\rm threshold}}                   \sim
\frac{ A_{\rm YMH}^{\star}}{d^{\star}}           =
\left( \frac{ A_{\rm YMH}^{\star}}{8 \pi^{2} /g^{2}}\right) \;
\left( \frac{ 2 }{d^{\star} \; M_{W}}               \right) \; \ES   \: ,
\label{eq:threshold}
\eeq
where we have used the definition
\[ \ES \equiv \pi\: \frac{M_{W}}{\alpha_{w}} \; ,\]
which is close to the true value $3.04 \, M_{W}/\alpha_{w}$
for the \sph ~energy at \mbox{$\lambda =0$.}
With the numerical estimates given in (\ref{eq:estimates}) we find,
not unexpectedly
perhaps, that the threshold in the parton \cme ~is of the order of
$\ES \sim 10\: {\rm TeV}$.
We intend to discuss the applications of \Istar ~in a separate publication.

\vspace{3\baselineskip}
\section*{Acknowledgements}
It is a pleasure to  thank
the experimental colleagues at NIKHEF-H for access to the Apollo
workstations, the staff of the Computer Group for technical
assistance, J. Smit for a valuable suggestion concerning the numerics
and M. Bonapart for help with the figures.
\newline This research  has been made possible in part by a fellowship of
the Royal Netherlands Academy of Arts and Sciences (KNAW).

\newpage
\section*{Appendix A : Numerical methods}
\appendix
\renewcommand{\theequation}{A.\arabic{equation}}
\setcounter{equation}{0}
We decribe in this appendix the algebraic and numerical calculations
for the results reported in sect. 5. These calculations are straightforward,
but cumbersome. Hence, we will give the main points only
and leave  many technical details out. In the first part of this appendix,
we review the algebraic calculation of the actiondensity
for our ansatz.
In the second part, we discuss the numerical solution of the variational
equations from this actiondensity.

\subsection*{A1 : Algebraic calculation}

The main algebraic calculation consists of two steps.
The first step is to insert the ansatz for the NCL into the constrained action.
We use axial coordinates
\begin{eqnarray}
x_{0}&\equiv&t                                                  \;,\nonumber\\
x_{3}&\equiv&z             \equiv r\:\cos\theta                 \;,\nonumber\\
x_{2}&\equiv&\rho\:\cos\phi\equiv r\:\sin\theta\:\cos\phi       \;,\nonumber\\
x_{1}&\equiv&\rho\:\sin\phi\equiv r\:\sin\theta\:\sin\phi       \;,
\label{eq:A1}
\end {eqnarray}
and make all distances dimensionless with $M_{W}^{-1}$.
The total constrained action (\ref{eq:A}) for the ansatz (\ref{eq:NCL})
takes the form
\beq
A = \frac{1}{g^{2}} \int_{-\infty}^{\infty} dt \: \int_{0}^{\infty} dr\:
r^{2}\:
    \int_{0}^{\pi} d\theta \sin\theta \: \int_{0}^{2\pi} d\phi \;\:a \;,
\label{eq:A2}
\eeq
with a rotationally invariant actiondensity
\beq
a = a(z,\rho,t) \; .
\label{eq:A3}
\eeq
As a matter of fact, it is already clear from the ansatz
(\ref{eq:NCL}, \ref{eq:U}) that a rotation of ($x_{1}$, $x_{2}$)
can be compensated by a global gauge transformation.

The second step is to perform two of the integrals in the action
(\ref{eq:A2}). The one over the azimuthal angle $\phi$ is trivial,
of course. The integral over the polar angle $\theta$ can also
be performed, since the functions $f,h$ and $g_{\pm}$ depend, by
construction, on $r$ and $t$ only. In addition, we have the
reflection symmetry $t \rightarrow -t$, so that the final expression
of the constrained action for our ansatz takes the form
\beq
A = \frac{8 \pi^{2}}{g^{2}} \int_{0}^{\infty} dt \: \int_{0}^{\infty}
    dr \: \frac{r^{2}}{\pi} \; \tilde{a}   \: .
\label{eq:A4}
\eeq
The averaged actiondensity $\tilde{a}=\tilde{a}(r,t)$
has the following structure
\begin{eqnarray}
 \tilde{a}         &=& \tilde{a}_{\rm YM} + \tilde{a}_{\rm H} +
                       \tilde{a}_{\rm C}  \; ,       \nonumber\\
           & &                                                \nonumber\\
 \tilde{a}_{\rm YM}&=& Z_{{\rm YM}20} \left(\partial_{t}f\right)^{2}+
               Z_{{\rm YM}02} \left(\partial_{r}f\right)^{2}+
               Z_{{\rm YM}11} \left(\partial_{t}f\partial_{r}f\right)+
               Z_{{\rm YM}00} \left(f(1-f)\right)^{2}      \;,\nonumber\\
           & &                                                \nonumber\\
 \tilde{a}_{\rm H} &=& 2          \left(\partial_{t}h\right)^{2}+
               2          \left(\partial_{r}h\right)^{2}+
               Z_{{\rm H}00}\left(h(1-f)       \right)^{2}+
               4 \frac{\lambda}{g^{2}} \left(h^{2}-1\right)^{2}\;, \nonumber\\
           & &                                                \nonumber\\
 \tilde{a}_{\rm C} &=& \kappa \left(
               Z_{{\rm C}20} \left(\partial_{t}f\right)^{2}+
               Z_{{\rm C}02} \left(\partial_{r}f\right)^{2}+
               Z_{{\rm C}11} \left(\partial_{t}f\partial_{r}f\right)
                             \right) \left(f(1-f)\right)^{2} \;,
\label{eq:A5}
\end{eqnarray}
where $\partial_{r}$ and $\partial_{t}$ denote partial derivatives with respect
to $r$ and $t$.
The eight coefficients $Z$ are complicated rational functions of the variables
$r$ and $t$, together with the functions $g_{\pm}(r,t)$ and their
various partial derivatives.
These coefficients depend also on the loop parameter $\omega$.
With some further effort, (\ref{eq:A5}) can be put in manifestly positive
definite form.
We have used the symbolic manipulation program FORM \cite{FORM}
for the algebraic calculation of the actiondensity (\ref{eq:A5}).

\subsection*{A2 : Numerical calculation}

We now have to solve numerically the variational equations from the action
(\ref{eq:A4}). We proceed in three steps.
The first step is to compactify the coordinates $r$ and $t$ to the
variables $x$ and $y$, respectively. Since we are interested in the
long-range behaviour of the Higgs field, we choose for $y(t)$
a rather slow dependence on $t$, specifically
\begin{eqnarray}
   y      &=& y_{c} + (1- y_{c}) \; \frac{\bar{t}-\bar{t}_{c}}
                                    {1+ | {\bar{t}-\bar{t}_{c}} | }\nonumber\\
   y_{c}  &\equiv&  \bar{t}_{c} /(1+ 2\: \bar{t}_{c})         \nonumber\\
   \bar{t}&\equiv&  t/t_{\rm scale}
\label{eq:A6}
\end{eqnarray}
and similarly for $x(r)$.  Furthermore, we let $t_{c}$ correspond to the
core position $t_{c} = D/2$ and set, typically,
$r_{c} = \rho_{\rm ~I}$ and $r_{\rm scale}=t_{\rm scale}= 2 \,\rho_{\rm ~I}$,
where
$\rho_{\rm ~I}= \rho_{\rm ~I}\:(\kappa)$ is the scale of the
instanton I, see sect. 5.1. It is straightforward to make these
changes of variables in the action
(\ref{eq:A4}),  and we write the result as
\beq
A = \frac{8 \pi^{2}}{g^{2}} \int_{0}^{1} dy \: \int_{0}^{1} dx \; \hat{a} \:,
\label{eq:A7}
\eeq
with the Jacobians absorbed into $\hat{a}=\hat{a}(x,y)$.
We look for the two functions $f(x,y)$ and $h(x,y)$ that
minimize this action, with mixed Dirichlet and Neumann boundary
conditions as indicated in fig. 7.

The second step is to discretize the integral (\ref{eq:A7}).
We use a rectangular grid for $x$ and $y$
\begin{eqnarray}
   x = i \Delta x  & \hspace{2cm} & i=0, \ldots,{\rm I} \nonumber\\
   y = j \Delta y  & \hspace{2cm} & j=0, \ldots,{\rm J}
\label{eq:A8}
\end{eqnarray}
and write for the functions at the lattice points
\beq
    f(i \Delta x, j \Delta y)  \rightarrow f_{i,j}   \; .
\label{eq:A9}
\eeq
Furthermore, we use central differences for the partial derivatives of
$f$ and $h$, for example
\beq
    \partial_{y}f(i \Delta x, j \Delta y)  \rightarrow
                \frac{ f_{i,j+1} -   f_{i,j-1}}{2 \Delta y} \; .
\label{eq:A10}
\eeq
The resulting discretized action is, however, numerically unstable :
minimization leads to functions $f_{i,j}$ and $h_{i,j}$  that take on
alternating values of approximately 0 and 1. We have chosen to
employ the following two countermeasures.
First, we ``double''  the boundary conditions at infinity, namely
\begin{eqnarray}
&    h_{i,{\rm J}} = h_{i,{\rm J}-1} = 1 & \nonumber \\
&    h_{{\rm I},j} = h_{{\rm I}-1,j} = 1 &\; ,
\label{eq:A11}
\end{eqnarray}
and similarly for $f$. Second, we ``smear''  the constraint term
$\hat{a}_{\rm C}(x,y)$ in the actiondensity. Specifically, we take
for the first term in $\hat{a}_{\rm C}$, cf. (\ref{eq:A5}),
\beq
 \kappa \; \hat{Z}_{{\rm C}20}
 \left( \frac{ f_{i,j+1} -   f_{i,j-1}}{2 \Delta y} \right)^{2}
        \left( \frac{ f_{i,j+1} +   f_{i,j-1}}{2} \right)^{2}
        \left( 1 - \frac{ f_{i,j+1} +   f_{i,j-1}}{2} \right)^{2} \; ,
\label{eq:A12}
\eeq
and similarly for the other terms.  In this way we end up
with a discretized actiondensity $\bar{a}_{i,j}$ at the gridpoint $(i,j)$.
For the total constrained action we have
\beq
A = \frac{8 \pi^{2}}{g^{2}} \;\sum_{i=0}^{\rm I}  \;
\sum_{j=0}^{\rm J} \; \Delta x \Delta y  \; w_{i,j} \; \bar{a}_{i,j}   \: ,
\label{eq:A13}
\eeq
with $w_{i,j}$ the weightfactors of, for example, the extended
trapezoidal rule.

The third, and final, step is to solve numerically the variational equations
for $f_{i,j}$ and $h_{i,j}$  coming from the discretized action
(\ref{eq:A13}).
These equations are highly non-linear, especially for the case
$\kappa=1$ we are interested in, and we need a method that can
handle this. We have succesfully employed the method of
non-linear overrelaxation (NLOR) \cite{Ames},
using grids of, typically, $25 \times 50$ points.
Our FORTRAN program starts with some trial functions for
$f_{i,j}$ and $h_{i,j}$,
which are then relaxed, first with NLOR, but in the end also with
some sweeps of {\em under\/}relaxation, if necessary.
In this way we obtain smooth configurations $\bar{f}$ and $\bar{h}$,
with a definite value for the constrained action.
The exact solution may even have a somewhat lower action than the one
obtained by relaxation.

\newpage
\section*{Appendix B : Non-contractible sphere}
\appendix
\renewcommand{\theequation}{B.\arabic{equation}}
\setcounter{equation}{0}

We consider in this appendix static, 3-dimensional configurations
of the gauge field $W$ and the Higgs field $\Phi$. The energy functional
for these fields is
\beq
 E_{\rm YMH}= \int_{{\rm R}^3} d^{3}x
 \left[ - \frac{1}{g^{2}}\; \frac{1}{2}\; {\rm Tr} \: W_{m n}^{2}
     + |D_{m} \Phi|^{2}
     + \lambda \left( |\Phi|^{2} - \frac{v^{2}}{2} \right)^{2} \: \right],
\label{eq:B1}
\eeq
with the same notation as in (\ref{eq:AYMH}). Henceforth, we set the
quartic Higgs coupling constant $\lambda=0$ and, for brevity, refer to
$E_{\rm YMH}$ as $E$.

It is well-known, by now, that a \ncl ~(NCL) of static configurations
leads to the existence of a static, but unstable, classical solution,
the sphaleron S \cite{DHN,KM}. It is not difficult to construct also
a non-contractible {\em sphere} (NCS) of static configurations \cite{K-Sstar}.
A mini-max procedure over this NCS suggests the possible existence
of a new sphaleron \Sstar, provided we exclude the case of two sphalerons
S at infinite separation. This loophole is closed if we are able
to construct a NCS, for which
\beq
{\rm max}\:  E_{{\rm NCS}}(\mu,\nu) < 2 \: E_{{\rm S}}\:  ,
\label{eq:B2}
\eeq
where $\mu$ and $\nu$ parametrize the position on the sphere.
Evidently, this discussion parallels the one for the new instanton \Istar
{}~in the main part of this paper. Moreover, the ansatz of sect. 4  can
easily be adapted to the present case. We will simply state the resulting
ansatz for the NCS and give our numerical results, which establish
the important inequality
(\ref{eq:B2}). These explicit results support the somewhat heuristic arguments
given in our previous paper \cite{K-Sstar}.

The NCS is parametrized by  the square $\mu,\nu \in [-\pi,+\pi]$, with the
boundary $ |\mu|=\pi$ or $|\nu|= \pi$ corresponding to the classical vacuum.
Writing $[\mu\nu] \equiv \max( |\mu|,|\nu|)$,
the configurations of the NCS are
\begin{eqnarray}
\pi/2 <[\mu\nu] \leq \pi & :
                         &W   =  0  \nonumber\\
                         &&\Phi=  \left( 1 - (1-h) \sin [\mu\nu]
                        \right)
                        \: \frac{v}{\sqrt{2}} \: \vectwo{0}{1}  \nonumber\\
                         &&  \nonumber\\
0 \leq[\mu\nu]\leq \pi/2 & :
                         &W   = -f\:  dU \: U^{-1}  \nonumber\\
                         &&\Phi= h  \: \frac{v}{\sqrt{2}} \: U \vectwo{0}{1},
\label{eq:B3}
\end{eqnarray}
with the following $SU(2)$ matrix $U$ for $\mu,\nu \in [-\pi/2,+\pi/2]$
\newpage
\begin{eqnarray}
U&=& \exp [ (\nu+\pi/2) i\sigma_{3}]                   \;
     \exp [ (\mu+\pi/2) \hat{y}_{-}\cdot i\vec{\sigma}]\;
     \exp [-(\nu+\pi/2) i\sigma_{3}] \cdot             \;\nonumber \\
 & & \exp [-(\mu+\pi/2) \hat{y}_{+}\cdot i\vec{\sigma}]
\label{eq:B4}
\end{eqnarray}
and notation
\begin{eqnarray}
x_{m\pm}&\equiv& (x_{1},\,x_{2},\,x_{3}\pm D/2)
         \equiv (       \rho\sin\phi,   \,\rho\cos\phi,\,z\pm D/2)\nonumber\\
y_{m\pm}&\equiv& (g_{\pm}\,\rho\sin\phi,\,g_{\pm}\,\rho\cos\phi,\,z\pm D/2) \;
{}.
\label{eq:B5}
\end{eqnarray}
The core distance $D=D(\nu)$ is given by
\beq
\begin{array}{rclcl}
\pi/2 <&|\nu|&\leq \pi    &\hspace{.5cm} :\hspace{.5cm}& D =0 \nonumber\\
0 \leq &|\nu|&\leq \pi/2  & : & D = d_{\rm max} \cos^{\delta}\nu  \; ,
\label{eq:B6}
\end{array}
\eeq
with $\delta > 0$ a free parameter. The axial functions
$f(\rho,z)$, $h(\rho,z)$ and $g_{\pm}(\rho,z)$
have the following boundary conditions and reflection symmetry
\begin{eqnarray}
\lim_{ |x| \rightarrow \infty} f,h,g_{\pm} &=& 1  \nonumber\\
f(0,\pm D/2)                               &=&h(0,\pm D/2)=0 \nonumber\\
f(\rho,z)                                  &=& f(\rho,-z)    \nonumber \\
h(\rho,z)                                  &=& h(\rho,-z)    \nonumber \\
g_{\pm}(\rho,z)                            &=& g_{\mp}(\rho,-z) \; .
\label{eq:B7}
\end{eqnarray}
For $g_{\pm}(\rho,z)$ we take again the functions (\ref{eq:gpm}),
with the same coefficients $\alpha$ and $\beta$, and where $x_{\pm}$ is now
defined by (\ref{eq:B5}). It remains to specify the two axial functions
$f$ and $h$. We will give two alternative constructions,
both of which lead to the desired inequality (\ref{eq:B2}).

The first construction for the functions $f$ and $h$ in the ansatz
(\ref{eq:B3}) parallels the procedure followed for the new instanton
in sect. 4, with the loop parameter  $\omega$ there corresponding to the
sphere parameter $\nu$ here.  The procedure is to solve the variational
equations for $f$ and $h$ at $\mu=0$ and $\nu=0$ or $\pi/2$, for
arbitrary values of the core distance $D$. We denote these solutions by
$\bar{f}(0,0,D)$ and $\bar{f}(0,\pi/2,D)$, and similarly for $\bar{h}$,
where the dependence on the spatial coordinates $\rho$ and $z$ is implicit.
In fig. 8 we show the corresponding energy values. For $\nu=0$ there is
a clear minimum
\footnote{
{}~The dip for the instanton case (fig. 4) is  more shallow,
one reason being the fact that the instanton functions  used,
$f(r,t)$ and $h(r,t)$, are not of the most general form,
which would be $f=f(\rho,z,t)$ and $h=h(\rho,z,t)$.
         }
at a core distance $d^{\star} \sim 7 \: M_{W}^{-1}$.
With these solutions $\bar{f}$ and $\bar{h}$ we can specify the functions
$f$ and $h$ in the ansatz (\ref{eq:B3})
\begin{eqnarray}
0 \leq |\nu| \leq \pi/2 &:& f=  \cos^{2}\nu \: \bar{f}(0,0    ,D) +
                         \sin^{2}\nu \: \bar{f}(0,\pi/2,D) \: \nonumber\\
                 & & h=  \cos^{2}\nu \: \bar{h}(0,0    ,D) +
                         \sin^{2}\nu \: \bar{h}(0,\pi/2,D) \: \nonumber\\
                 & & D=  d_{\rm max} \: \cos^{\delta}\nu      \nonumber\\
\pi/2 < |\nu| \leq \pi  &:& h=  \bar{h}(0,\pi/2,0) \; .
\label{eq:B8}
\end{eqnarray}
Choosing $d_{\rm max} = d^{\star}$ and the parameter $\delta$
sufficiently small, we have verified that the energy stays
everywhere below $2 \: \ES$. In particular, the energy over slices of the NCS
at
constant values of $\nu$ have their maximum at $\mu=0$ and
drop to zero monotonically for $\mu \rightarrow \pm \pi$. The energy
surface, however, is very steep for $\nu \sim \pm \pi/2$, and we prefer to
show it for a somewhat different construction.

This second, alternative construction for the functions $f$ and $h$
in the ansatz (\ref{eq:B3}) is as follows.
The procedure is to solve the variational equations for $f$ and $h$
at $\mu=0$ for the {\em whole} range of values $\nu \in [-\pi/2,+\pi/2]$,
varying the core distance simultaneously
$ D=  d_{\rm max} \: \cos^{\delta}\nu  $.
Actually, we do not need the complete range for $\nu$,
only the interval $[-\nu_{\rm min},+\nu_{\rm min}]$, so that
$D$ runs from $d_{\rm max}$ to  some smaller value $d_{\rm min}$
($ \equiv  d_{\rm max} \: \cos^{\delta}\nu_{\rm min}$).
We denote these solutions by $\bar{f}(0,\nu,D;\rho,z)$ and
$\bar{h}(0,\nu,D;\rho,z)$, where now the dependence on the
spatial coordinates $\rho$ and $z$ is explicit. For
$\nu \in [-\pi/2,-\nu_{\rm min}] \cup [+\nu_{\rm min},+\pi/2]$
we use a simple deformation of the functions
$\bar{f}$ and $\bar{h}$ at $\nu =\nu_{\rm min}$, in order to move the
zeros of $f$ and $h$  appropriately.
We denote the three segements of the NCS by
\[
\begin{array}{lcrcl}
{\rm III} &:& \hspace{1cm} 0 \leq          & |\nu| &\leq \nu_{\rm min} \\
{\rm II}  &:&              \nu_{\rm min} < & |\nu| &\leq \pi/2         \\
{\rm I}   &:&              \pi/2 \leq      & |\nu| &\leq \pi
\end{array}
\]
and take for the functions $f$ and $h$ in the ansatz (\ref{eq:B3})
\begin{eqnarray}
{\rm III} &:&   f= \: \bar{f}(0,\nu,D;\rho,z)       \nonumber\\
          & &   h= \: \bar{h}(0,\nu,D;\rho,z)       \nonumber\\
          & &   D=    d_{\rm max}\:\cos^{\delta}\nu \nonumber\\
       &     &                                      \nonumber\\
{\rm II~}  &:&
f=\frac{x_{+}^{2}\,
   \bar{f}(0,\nu_{\rm min},d_{\rm min};\rho,z+d_{\rm min}/2-D/2)+ x_{-}^{2}\,
   \bar{f}(0,\nu_{\rm min},d_{\rm min};\rho,z-d_{\rm min}/2+D/2)}
   {x_{+}^{2}+ x_{-}^{2} }               \nonumber\\
&&h=\frac{x_{+}\,
   \bar{h}(0,\nu_{\rm min},d_{\rm min};\rho,z+d_{\rm min}/2-D/2)+ x_{-}\,
   \bar{h}(0,\nu_{\rm min},d_{\rm min};\rho,z-d_{\rm min}/2+D/2)}
   {x_{+}+ x_{-} }                     \nonumber\\
       &     &                                 \nonumber\\
&&x_{\pm}\:\:\equiv \left[ \rho^{2} + (z \pm D/2)^{2}\right]^{1/2}  \nonumber\\
&&D\:\:\:\:=  d_{\rm max}\:\cos^{\delta}\nu            \nonumber\\
&&d_{\rm min} =  D(\nu_{\rm min})         \nonumber\\
       &     &                                 \nonumber\\
{\rm I~~}   &:&h=\frac{
     \bar{h}(0,\nu_{\rm min},d_{\rm min};\rho,z+d_{\rm min}/2)+
     \bar{h}(0,\nu_{\rm min},d_{\rm min};\rho,z-d_{\rm min}/2)}{2}
                                        \; .
\label{eq:B9}
\end{eqnarray}
This construction leads to a smooth energy surface over the NCS,
shown in fig. 9 for one particular set of parameters.
The energy surface over the NCS
is, by construction, invariant under $(\mu,\nu) \rightarrow (-\mu,-\nu)$.
Again, the maximum energy over the NCS is reached at
$\mu=\nu=0$ and its value ( $1.94 \: \ES$, for the parameters chosen in fig. 9)
obeys the inequality (\ref{eq:B3}).

To summarize, we have constructed in $SU(2)$ \YMHth ~\ncss
{}~of static configurations, with energies everywhere below $2\: \ES$.
This suggests the existence of a new sphaleron \Sstar.
Furthermore, we have, for the case of vanishing Higgs mass,
an approximation of that solution
from the optimal maximum configuration on the NCS, with a
core distance and energy given by (see fig. 8)
\begin{eqnarray}
       d^{\star}  &\sim& 7 \: M_{W}^{-1} \nonumber\\
       E^{\star}  &\sim&  1.92 \: \ES  \; ,
\label{eq:B10}
\end{eqnarray}
where $E_{\rm S}$ takes the numerical value $3.04 \, M_{W}/\alpha_{w}$.
In fig. 10 we show the energy density of this configuration.
Finally, we note that the 3-dimensional configuration (\ref{eq:B3})
at $\mu=\nu=0$ is essentially equivalent to the slice $x_{3}=0$ of the
instanton configuration (\ref{eq:NCL}, \ref{eq:U}) at $\omega=0$.
It is in this sense that the new sphaleron \Sstar ~corresponds to a constant
time slice of the new instanton \Istar.

\newpage
\vspace*{4\baselineskip}
\begin{table}
\newcommand {\bk} {\kappa}
\begin{center}
\begin{tabular}{|l||c|c|c||c|c|c||c|c|c|}
\hline
& \multicolumn{3}{c||}{$M_{H}/M_{W}=0$}
& \multicolumn{3}{c||}{$M_{H}/M_{W}=1$}
& \multicolumn{3}{c|}{$M_{H}/M_{W}=10$}\\
      &$\rho$  & $A$    &$A_{\rm YMH}$& $\rho$ &  $A$   &$A_{\rm YMH}$
				      & $\rho$ &  $A$   &$A_{\rm YMH}$
								    \\
\hline
$\bk=10^{-4}$ & $0.39$ & $1.10$ & $1.07$      & $0.39$ & $1.10$ & $1.07$
				      & $0.37$ & $1.11$ & $1.08$    \\
$\bk=10^{-3}$ & $0.58$ & $1.19$ & $1.13$      & $0.57$ & $1.20$ & $1.14$
				      & $0.54$ & $1.24$ & $1.16$    \\
$\bk=10^{-2}$ & $0.87$ & $1.39$ & $1.27$      & $0.85$ & $1.41$ & $1.28$
				      & $0.80$ & $1.51$ & $1.34$    \\
$\bk=10^{-1}$ & $1.30$ & $1.79$ & $1.55$      & $1.26$ & $1.86$ & $1.58$
				      & $1.19$ & $2.07$ & $1.73$    \\
$\bk=10^{+0}$ & $1.93$ & $2.61$ & $2.11$      & $1.85$ & $2.80$ & $2.21$
				      & $1.75$ & $3.26$ & $2.52$    \\
$\bk=10^{+1}$ & $2.85$ & $4.30$ & $3.26$      & $2.69$ & $4.88$ & $3.56$
				      & $2.58$ & $5.77$ & $4.21$    \\
\hline
\end{tabular}
\end{center}
\caption[]{ \protect
Numerical results for the BPSTH instanton (\protect \ref{eq:ansatzI})
in the constrained \YMHth ~(\protect \ref{eq:A}).
The classical theory has two dimensionless coupling constants $\kappa$ and
$\lambda/g^{2}$, which control, respectively, the strength of
the constraint term in the action
and the mass ratio of the Higgs scalar $H$  and vector bosons $W$.
The scale $\rho$ of the instanton is given in units of $M_{W}^{-1}$,
the constrained action $A$ and the standard Yang-Mills-Higgs action
$A_{\rm YMH}$ in units of $8 \pi^{2}/g^{2}$.
}
\end{table}

\newpage
\section*{Figure captions }
\par
{\bf Fig. 1 :} {\bf (a)} Sketch of the constrained action $A$ as a function
of the core distance $D$, for the case of repulsive ($\omega=0$) or
attractive ($\omega=\pi/2$) Yang-Mills interactions.
The Higgs mass vanishes and for large distances there is attraction between
the cores. The horizontal
dashed line corresponds to twice the action of the BPSTH instanton.
Also indicated are the minimum value $A^{\star}$ on the $\omega=0$ curve
and the corresponding distance \dstar.
\newline
{\bf (b)} Alternative behaviour of $A$ vs. $D$.
\vspace{1\baselineskip}
\newline
{\bf Fig. 2 :}  Yang-Mills-Higgs action $A_{\rm YMH}$
(in units of $8\pi^{2}/g^{2}$)
of the numerical solution of the variational equations for the BPSTH
ansatz (\ref{eq:ansatzI}), as a function of the scale $\rho$
(in units of $M_{W}^{-1}$).
The quartic Higgs coupling constant $\lambda$ vanishes.
The dashed curve represents the analytical
result (\ref{eq:AYMHI}), valid for small values of $\rho$.
\vspace{1\baselineskip}
\newline
{\bf Fig. 3 :} Constrained action $A$ (normalized to its asymptotic value)
of the numerical solution of the variational equations for the
configurations of the \ncl
{}~(\ref{eq:NCL}), as a function of the core distance $D$
(in units of $M_{W}^{-1}$).
Closed and open symbols correspond to, respectively, repulsive ($\omega=0$) and
attractive ($\omega=\pi/2$) Yang-Mills interactions.
The coupling constants $\lambda$ and $\kappa$ in the constrained action
(\ref{eq:A}) have the values 0 and 1, respectively.
\vspace{1\baselineskip}
\newline
{\bf Fig. 4 :} Same as fig. 3, but with an expanded scale for the
               constrained action $A$.
\vspace{1\baselineskip}
\newline
{\bf Fig. 5 :} Same as fig. 3, but for a smaller value of the constraint
coupling constant $\kappa$.
\vspace{1\baselineskip}
\newline
{\bf Fig. 6 :} Actiondensity $\tilde{a}_{\rm YMH}(r,t)$ and Pontryagin-density
$\tilde{q}(r,t)$  for the optimal maximum configuration of
the \ncl ~($\omega=0,\:D=10 \: M_{W}^{-1},\:\lambda=0,\:\kappa=1$), see fig. 4.
Both densities have an arbitrary normalization of their maximum to 100.
The coordinates $r$ and $t$ are in units of $M_{W}^{-1}$.
The  complete configuration is obtained by reflection $t \rightarrow -t\:$,
under which the actiondensity is invariant, but the Pontryagin-density
changes sign.
\vspace{1\baselineskip}
\newline
{\bf Fig. 7 :} Boundary conditions for the functions $f(x,y)$ and
$h(x,y)$, defined over the unit square. The Dirichlet   boundary
conditions are 0 at the core ($x=0,\:y=y_{c}$) and 1 at spatial
infinity ($x=1$ or $y=1$). Neumann boundary conditions, for
$x=0$ or $y=0$, are indicated by N.
\vspace{1\baselineskip}
\newline
{\bf Fig. 8 :} Energy $E$ (normalized to its asymptotic value)
of the numerical solution of the variational equations for the
configurations of the \ncs ~(\ref{eq:B3}),
as a function of the core distance $D$ (in units of $M_{W}^{-1}$).
Closed and open symbols correspond to different parameters of the ansatz,
respectively ($\mu=0,\:\nu=0$) and ($\mu=0,\:\nu=\pi/2$).
The quartic Higgs coupling constant $\lambda$ vanishes.
\vspace{1\baselineskip}
\newline
{\bf Fig. 9 :} Energy $E$ (with arbitrary normalization) over the
non-contractible sphere (\ref{eq:B3}, \ref{eq:B9}).
The energy surface over the whole sphere is obtained by reflection symmetry
$(\mu,\,\nu) \rightarrow (-\mu,\, -\nu)$. The parameters for these results are
$d_{\rm max} = 10 \; M_{W}^{-1}$, $\nu_{\rm min}= 7 \pi/16$ and $\delta=1$.
\vspace{1\baselineskip}
\newline
{\bf Fig. 10:} Energydensity $e(\rho,z)$ (with arbitrary normalization)
for the optimal maximum configuration of
the \ncs ~($\mu=\nu=0,\: D=7 \: M_{W}^{-1},\:\lambda=0$), see fig. 8.
The coordinates $\rho$ and $z$ are in units of $M_{W}^{-1}$.
The  complete configuration is obtained by reflection $z \rightarrow -z$.

\end{document}